\documentclass[%
 reprint,
 amsmath,amssymb,
 aps,
]{revtex4-2}

\usepackage[english]{babel}
\usepackage{xcolor} 
\usepackage[normalem]{ulem}
\usepackage{ifthen}

\usepackage{mathtools}
\usepackage{graphicx}
\usepackage{dcolumn}
\usepackage{bm}
\usepackage{hyperref}

\usepackage[
scale=0.7, marginratio={1:1, 2:3}, ignoreall,
total={6.5in,8.75in}, top=1.2in, left=0.9in, includefoot,
]{geometry}

\newif\ifuseSRversion
\useSRversiontrue   

\newif\ifclean
\cleantrue         

\ifuseSRversion
  \ifclean
    \newcommand{\added}[1]{#1}
    \newcommand{\deleted}[1]{}
    
    \newcommand{\reviewnote}[1]{}
  \else
    \newcommand{\added}[1]{\textcolor{blue}{#1}}
    \newcommand{\deleted}[1]{\textcolor{red}{[\,#1\,]}}
    
    \newcommand{\reviewnote}[1]{\textcolor{teal}{\textsf{\footnotesize\ [SR: #1]}}}
  \fi
\else
  \newcommand{\added}[1]{#1}
  \newcommand{\deleted}[1]{}
  
  \newcommand{\reviewnote}[1]{}
\fi

\begin{document}

\preprint{PRL/123-QED}

\title{A nanoscale magnetic spectrum analyzer based on qubit dressed states}

\author{Jan Rueschkamp$^{1,4*}$, Shantam Ravan$^{1,5}$, Daniel Fernandez$^{1}$, Nazar Delegan$^{6,7,8}$,F. Joseph Heremans$^{6,7,8}$, David D. Awschalom$^{6,7,8,9}$, Ronald L. Walsworth$^{5}$, Nikola Maksimovic$^{1,3*}$, Amir Yacoby$^{1,2}$}

\email{jrueschkamp@fas.harvard.edu
\\nmak@bu.edu
\\yacoby@g.harvard.edu}

\affiliation{
$^1$Department of Physics, Harvard University, Cambridge, MA, USA \\
$^2$John A. Paulson School of Engineering and Applied Sciences, Harvard University, Cambridge, MA, USA \\
$^3$Department of Physics, Boston University, Boston, MA, USA
\\
$^4$Department of Physics, Eidgenössische Technische Hochschule Zürich, Zürich, Switzerland
\\
$^5$Departmet of Physics, University of Maryland, College Park, MD, USA
\\
$^6$Pritzker School of Molecular Engineering, University of Chicago, Chicago, Illinois 60637, USA
\\
$^7$Q-NEXT, Argonne National Laboratory, Lemont, Illinois 60439, USA
\\
$^8$Materials Science Division, Argonne National Laboratory, Lemont, Illinois 60439, USA
\\
$^9$Department of Physics, University of Chicago, Chicago, Illinois 60637, USA
}

\date{\today}

\begin{abstract}
Magnetic field fluctuations on nanometer length scales manifest in a diverse range of phenomena --- electron and spin dynamics in materials and devices, quantum many-body systems, and molecular chemistry. Measuring these phenomena requires sensors with a challenging combination of broad spectral bandwidth, high sensitivity, and nanoscale spatial resolution. Nitrogen-vacancy (NV) centers, atom-like quantum sensors in diamond, possess the requisite sensitivity and nanoscale sensing volume, but are typically limited in bandwidth by the practical speed of the applied quantum control sequence. Here, we overcome this limitation by exposing the NV qubit to a microwave dressing field during a dynamical decoupling sequence, which both amplifies and frequency-mixes target signals at arbitrary frequencies into the detection band of the dynamical decoupling protocol. We demonstrate this approach by using NV centers to detect both coherent and noisy nanoscale spin wave dynamics in a magnetic yttrium-iron-garnet (YIG) thin film over a broad frequency range. Our technique generalizes to other qubit platforms, providing a versatile framework for nanoscale spectroscopy across diverse physical and chemical systems.
\end{abstract}

\maketitle

\section{Introduction}
A variety of phenomena in physics, chemistry, and quantum technology manifest as magnetic fluctuations on nanometer length scales. Examples include spin dynamics in magnets~\cite{xue_magnon_2025} and spin liquids~\cite{xiao_signatures_2025,lee_proposal_2023}, electrical current dynamics in superconductors~\cite{curtis_probing_2024, liu_quantum_2025} and hydrodynamic metals~\cite{halperin_hydrodynamic_1969}, spin dynamics and correlations in small-scale chemistry~\cite{aslam_quantum_2023,glenn_high-resolution_2018}, and noise in quantum device hardware~\cite{bar-gill_suppression_2012}. Fluctuation phenomena can even be used to extract entanglement properties of a many-body quantum system~\cite{hauke_measuring_2016}.

For the purpose of detecting these phenomena, quantum sensors based on atomic-scale color centers in semiconductors, such as the nitrogen-vacancy (NV) center in diamond, are a promising platform that sets new benchmarks in magnetic field sensitivity and spatial resolution~\cite{barry_sensitivity_2020}. These solid-state spin qubit sensors measure radio-frequency (RF) magnetic fields using dynamical decoupling (DD) protocols~\cite{hahn_spin_1950,de_lange_universal_2010}, a technique that manipulates the spin state of the qubit by using microwave (MW) control pulses, allowing for both homodyne~\cite{taylor_high-sensitivity_2008,meinel_high-resolution_2023} and heterodyne~\cite{glenn_high-resolution_2018} detection of coherent signals, as well as noise spectroscopy~\cite{xue_magnon_2025}. The latter, in particular, is vital for the investigation of many physical phenomena where spectral responses are often broadband~\cite{cronenberger_atomic-like_2015}. While DD noise magnetometry using spin-qubits provides an unparalleled combination of sensitivity, spatial resolution, and practicality, it is limited in frequency bandwidth and range by the speed of the DD qubit control sequence as defined by the Rabi rate -- proportional to the amplitude of the pulsed control field. In NV diamond, for example, DD noise sensing is limited to $\sim$100 MHz~\cite{jung_impedance-tuned_2025,fuchs_gigahertz_2009}. This bandwidth/range limitation is coincidentally similar in other nanoscale RF magnetic probes like RF SQUIDs~\cite{foroughi_micro-squid_2018}.

In this work, we introduce a method for spectral analysis of RF noise, based on well-established DD protocols, that extends the accessible frequency range beyond the Rabi limit by leveraging qubit dressed states~\cite{wang_sensing_2022}. We demonstrate this dressed state DD technique using small NV ensembles to detect both artificial test signals and the spin wave dynamics of a magnetic thin film of Yttrium Iron Garnet (YIG). From the resulting measured RF spectra, we characterize magnetic excitations in the YIG film under conditions that are inaccessible with other quantum sensing protocols, highlighting the value of the dressed state DD technique in practical sensing applications.

\section{results}

\subsection{Demonstration on a test signal}

Our technique relies on parametrically mixing a weak RF noise signal with a strong MW probe field. This mixing is enabled by the quadratic scaling of the AC Stark effect on the total AC magnetic field amplitude. This quadratic dependence gives rise to a measurable cross term -- a term proportional to the product of the signal and probe field. Embedding this mixing within a DD protocol allows us to amplify and measure a target noise signal at arbitrary frequency. Note that the technique presented here complements recent work using quantum frequency mixing to extend the frequency range of high-spectral-resolution NV sensing of coherent RF signals~\cite{wang_sensing_2022,yin_high-resolution_2025}.

\begin{figure}
    \centering
    \includegraphics[width=1\linewidth]{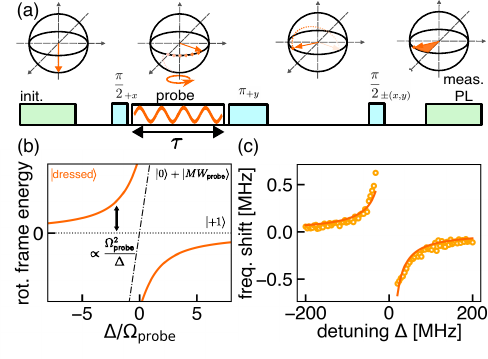}
    \caption{Measurement of qubit dressed states using a spin echo protocol. (a) Dressed state spin echo sequence with a probe field during one arm of length $\tau$, which acts as an effective B field on the qubit (see supplemental section \ref{chap:derivation_Stark}). Bloch sphere representation (above) shows effect of pulse sequence on qubit spin. (b) Calculated dressed state energies (within the rotating frame) in comparison to the qubit bare states as a function of detuning, $\Delta$, of the probe field from the bare resonance relative to the probe field Rabi rate, $\Omega_{\text{probe}}$. The black dotted line describes the bare qubit excited state in the rotating frame, while the dashed line gives the summed energy of the bare qubit ground state and the MW probe field photon. The orange lines are the dressed states, which diverge from the bare states due to the AC Stark effect with a quadratic dependency on the probe field strength, $\Omega_{\text{probe}}$. (c) NV spin echo measurement of the dressed state frequency shift induced by the probe field as a function of detuning $\Delta$, and a fit of the expected $1/\Delta$ dependency for $\tau = 1500$ ns. From the fit we obtain $\Omega_\text{probe} = 2\pi \times (3.70 \pm 0.04)$MHz.}
    \label{fig:1}
\end{figure}

Before introducing the signal of interest, we first establish the origin of this nonlinearity, which arises from the AC Stark shift: an off-resonant MW probe field at frequency $f_\text{probe}$ hybridizes the qubit eigenstates into dressed states whose energy splitting is shifted relative to the bare splitting by an amount set by the probe field strength and detuning~\cite{wei_strongly_1997}. We measure the AC Stark shift using a standard spin echo DD sequence (Fig.~\ref{fig:1}(a)), applying the probe field during one arm of the echo. The resulting energy shift acts as an effective static magnetic field along the qubit quantization axis, causing the qubit to accumulate a phase on the Bloch sphere equator, which we read out via quantum state tomography (see supplemental section~\ref{chap:measurement_sequence}).

We define the detuning $\Delta = f_\text{probe} - f_\text{res}$, where $f_\text{res}$ is the bare resonance frequency of the qubit in the absence of any applied MW drive fields, and the probe field strength is $\Omega_\text{probe} = \gamma B_\text{probe}/\sqrt{2}$, where $\gamma$ is the gyromagnetic ratio, and $B_{\text{probe}}$ is the probe field amplitude.

In a frame rotating at $f_\text{res}$, the dressed states (see supplemental section~\ref{chap:derivation_Stark}) have energies: 

\begin{equation}
E_\pm = \pm \frac{h}{2}\sqrt{\Delta^2+\Omega_\text{probe}^2} \approx \pm\frac{h}{2}( \Delta +\frac{\Omega_\text{probe}^2}{2\Delta}),
\label{eq:def_eigenenergies}
\end{equation}
where the last step assumes operation in the high detuning regime $\Delta\gg\Omega_\text{probe}$. The energy shift relative to the bare splitting is $\delta E = (E_+-E_-) - h \Delta= \added{h}\frac{\Omega^2_\text{probe}}{2\Delta}$, which scales quadratically with the Rabi rate, i.e., with the probe field amplitude (Fig.~\ref{fig:1}(b)). The qubit phase accumulated during an echo arm of length $\tau$ is:

\begin{equation}
\phi = \frac{1}{h}\int_0^\tau dt\, \delta E
=-\frac{\Omega^2_\text{probe}}{2\Delta} \tau.
\label{eq:coherent_phase}
\end{equation}

We implement this protocol experimentally using a room-temperature diamond pillar containing a small nanoscale ensemble of NV centers (see supplemental section~\ref{chap:pillar_sample}) that is subject to a 0.3 mT external magnetic field along the pillar axis to split the $|-1\rangle$ and $|+1\rangle$ electronic spin transitions in the NV ground state manifold. MW signals are delivered via a wire bond near the pillar. The measurement qubit is encoded in the $|0\rangle$ and $|+1\rangle$ states and optically initialized and read out using standard NV techniques~\cite{maze_nanoscale_2008, barry_sensitivity_2020}.

Fig.~\ref{fig:1}(c) shows the measured dressed state frequency shift $\delta f = \frac{\phi}{2\pi\tau}$ as a function of the probe/resonance detuning; applying a $1/\Delta$ fit to the data using Eq.~\ref{eq:coherent_phase}, we extract $\Omega_\text{probe} = 2\pi \times (3.70 \pm 0.04)$ MHz.

If the spin qubit is also subject to a noisy RF signal, the total Rabi rate becomes $\Omega_\text{eff}= \Omega_\text{probe}+\Omega_\text{signal}$. Because the AC Stark shift depends on $\Omega_\text{eff}^2$ (Eq.~\ref{eq:def_eigenenergies}), expanding the square produces the terms $\Omega_\text{probe}^2$ and $\Omega_\text{signal}^2$, as well as a cross term $\propto \Omega_\text{probe}\cdot\Omega_\text{signal}$. Intuitively, the cross term can be used as a parametric mixer because its amplitude scales with the probe field strength~\cite{wang_sensing_2022}. Since the noisy signal field is weak and far off resonance, the $\Omega_\text{signal}^2$ term is negligible (see supplemental section~\ref{chap:spectrum_analyzer}).

\begin{figure}
    \centering
    \includegraphics[width=1\linewidth]{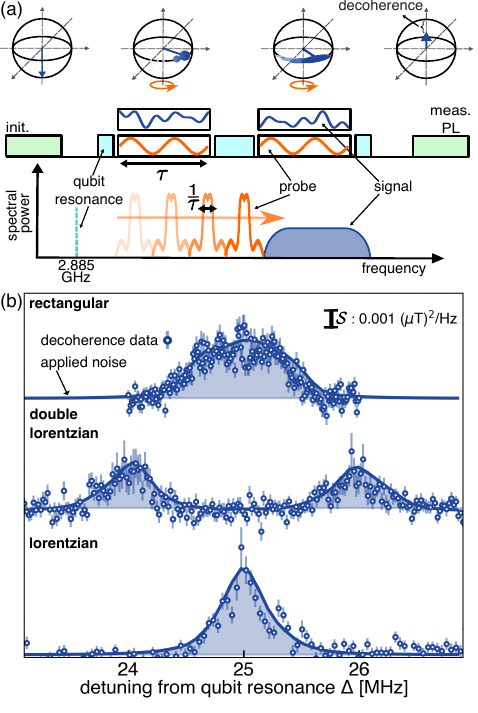}
    \caption{Measurement of broadband noise spectra using qubit dressed states. (a) Dressed state spin echo sequence with probe field and noisy RF signal during both arms of length $\tau$. The noise mixes with the probe field via the quadratic dependence of the qubit AC Stark shift on the total field amplitude. This mixing effectively imprints the noise onto the probe field, producing a fluctuating effective Rabi rate that induces decoherence in the qubit’s $x$–$y$ plane. Sweeping the spin echo sequence filter function, which is set by the probe field frequency $f_\text{probe}$, and repeating spin echo measurements allows determination of $S$, the noise spectral density of the RF signal field. Bloch sphere representation (upper panel) shows effect of pulse sequence and applied fields on the qubit spin. Schematic spectrum (lower panel) illustrates relative frequency locations and bandwidths of the qubit bare resonance, swept spin echo filter function, and noise signal. (b) Noise spectral density from NV decoherence measurements for three noise spectra. Solid lines represent independent measurements obtained using a conventional spectrum analyzer. The scale of $S$ is indicated at the top of the panel.}
    \label{fig:2}
\end{figure}

Because the noisy signal field is broadband, embedding the probe field in both arms of a DD sequence (Fig.~\ref{fig:2}a) selects a specific spectral component of the cross term. Moreover, the phase-evolution symmetry of this dressed state  DD sequence cancels the deterministic AC Stark Shift term $\propto \Omega_\text{probe}^{2}$. Computing the resulting qubit phase variance $\langle\phi^2\rangle$ (see supplemental section~\ref{chap:spectrum_analyzer}) yields:
\begin{eqnarray}
    \label{eq:var_def_convolution}
    \langle \phi^2\rangle(f_\text{probe}) &=& 2\frac{\Omega_{\text{probe}}^2}{\Delta^2}\int_{-\infty}^{\infty} df\,
    \mathcal{S}(f) W_\tau(f - f_\text{probe})\nonumber \\
    &=&2 \frac{\Omega_{\text{probe}}^2}{\Delta^2}(\mathcal{S}* W_\tau)(f_\text{probe})
\end{eqnarray}
where $\mathcal S$ is the noise spectral density of the signal field and $W_\tau$ is the filter function of the spin sequence, centered at the probe field frequency $f_{\text{probe}}$ and with bandwidth $\approx 1/\tau$. The operation $*$ denotes a convolution. Intuitively, the dressed state DD sequence provides a tunable RF noise detector and amplifier, where $f_\text{probe}$ scans the filter across the RF noise spectrum, and the probe field amplifies the target noise signal (Fig.~\ref{fig:2}a, lower panel). 

We determine the phase variance via the NV spin coherence magnitude $\Lambda$, given by the total photoluminescence (PL) contrast of the NV spin at the end of repeated spin echo measurements with a fixed evolution time (see supplemental section~\ref{chap:measurement_sequence})~\cite{machado_quantum_2023}. Normalization is given by a probe-free reference measurement, $\Lambda_\text{ref}$, to remove the effect of intrinsic decoherence: $\langle\phi^2\rangle(f_\text{probe}) =-\frac{1}{2} \ln\left(\frac{\Lambda(f_\text{probe})}{\Lambda_\text{ref}}\right)$. When $\mathcal S$ is much broader than $W_\tau$, 
\begin{equation}
\begin{aligned}
    \langle\phi^2\rangle(f_{\text{probe}})
    &=\mathcal S(f_{\text{probe}}) \,
2\tau
\left(\frac{2\gamma B_\text{probe}}{\Delta}\right)^2.
\end{aligned}
\label{eq:calc_spectral_density}
\end{equation}

As a proof of principle experiment using the small NV ensemble in the diamond pillar, we apply the measurement sequence in Fig.~\ref{fig:2}a to probe broadband noise spectra centered at $\sim2.9$ GHz that are generated as described in supplemental section~\ref{chap:Noise_creation}; this frequency regime is inaccessible to conventional NV DD measurement protocols due to the limited Rabi rate for driving NV electronic spins \cite{yin_high-resolution_2025}. Fig.~\ref{fig:2}(b) compares the NV data for three classes of broadband noise spectra with independent measurements from a commercial spectrum analyzer, showing excellent agreement.

\begin{figure}
    \centering
    \includegraphics[width=1\linewidth]{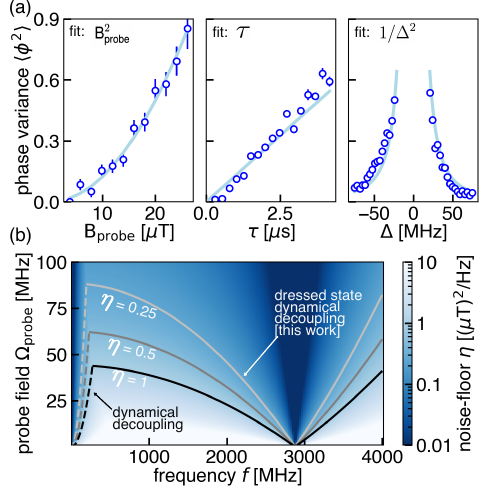}
    \caption{Scaling of dressed state DD RF noise measurements on experimental parameters. (a) Dependence of the experimentally determined NV spin phase variance $\langle\phi^2\rangle$ on probe field amplitude, $B_{\text{probe}}$, measurement time, $\tau$, and probe field detuning from the NV resonance, $\Delta$. The data is well-fitted by the scalings predicted by our model of this measurement scheme (Eq.~\ref{eq:calc_spectral_density}). (b) Estimate of noise floor, i.e. minimal detectable RF noise spectral density, as a function of Rabi rate $\Omega_{\text{probe}}$ and noise signal frequency $f$, obtained by extrapolating the noise floor in our experiment for one second of integration where the Rabi rate $\approx$ 2.5 MHz.}
    \label{fig:3}
\end{figure}

For fixed signal noise spectral density, $S$, Eq.~\ref{eq:calc_spectral_density} predicts that $\langle\phi^2\rangle$ scales with $B^2_\text{probe}$, $\tau$, and $1/\Delta^2$, all of which are borne out in the experimental data (Fig.~\ref{fig:3}(a)). The small deviation in linear $\tau$ dependence for $\tau < 700$ ns can be attributed to the fact that the noise bandwidth approaches the bandwidth of the spin echo measurement (see supplemental section \ref{chap:scalings}). 
 
Crucially, the actual measured decoherence depends on the probe field strength, which can be increased arbitrarily to amplify a target noise signal.
The measurement here is done at $\Omega_\text{probe} \sim 2.5$~MHz. In practice, NV Rabi rates $\sim 100$~MHz are experimentally attainable~\cite{jung_impedance-tuned_2025,fuchs_gigahertz_2009}, enabling the detection of noisy RF signals across a wide frequency range. Quantitatively, we define the noise floor $\eta$ through the uncertainty of a data point in Fig.~\ref{fig:2}(b) normalized to one second of integration time, $\eta \equiv \delta\mathcal{S}(f_{\text{res}}+26.5~\text{MHz}) = 5900~(\text{nT})^2/\text{Hz}$. Extrapolating this noise floor to higher probe field strengths using Eq.~\ref{eq:calc_spectral_density} reveals that noise signals of $100~(\mu\text{T})^2/\text{Hz}$ over many GHz in frequency could feasibly be amplified to a detectable threshold (Fig.~\ref{fig:3}(b)). The non-linearity of the scaling below 2 GHz is due to the counter rotating component of the AC Stark shift that is otherwise dropped in the rotating wave approximation (see supplemental section \ref{chap:derivation_Stark}). 

Fig.~\ref{fig:3}(b) further compares the noise floor of our dressed state DD to that of conventional DD noise magnetometry under comparable conditions. Conventional DD noise magnetometry is analyzed via the following relationship between $\left< \phi^{2} \right>$ and the noise floor $\eta$. In the limit of target noisy signal frequency much greater than DD sensing frequency, the decoherence signal in DD is given by $\left< \phi^{2} \right> = \gamma^{2} W_{\tau}(f)*\tau_{\text{eff}}*\eta$, where $W_{\tau}(f) \sim 1/f^{2}$, and the effective evolution time, $\tau_{\text{eff}}$, is the total evolution time minus the time spent on $\pi$-pulses ($\sim 1/\Omega_{\text{Rabi}}$). In Fig.~\ref{fig:3}(b), the Rabi rate in conventional DD is set equal to the probe field strength in dressed state DD for the purpose of comparing the two protocols.

\subsection{Measurements on a YIG thin film}

We next apply the dressed state DD protocol to NV centers in proximity to a thin film of yttrium iron garnet (YIG), extracting both the magnon noise spectrum and the group velocity by sweeping the probe field frequency and making repeated spin echo measurements, as with the RF noise signal demonstration described above. These experiments are performed with a room temperature apparatus (see supplemental Fig.~\ref{fig:Yig_sample_zoom} for details).

\begin{figure}
    \centering
    \includegraphics[width=1\linewidth]{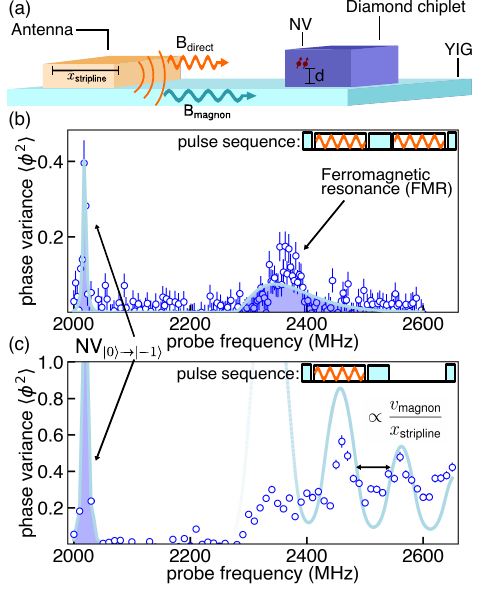}
    \caption{Probing the spin wave properties of YIG using dressed state DD. (a) Experimental setup consists of a gold stripline antenna that produces a MW probe field to coherently drives the electronic spins of an NV ensemble in two ways: directly and by launching magnons into the thin YIG film beneath. The NV spins are also affected by intrinsic thermal magnon MW noise produced by the YIG.  (b) Dressed state spin echo sequence, for which the experimentally determined phase variance $\langle\phi^{2}\rangle$ of the NV sensor as a function of probe field frequency resolves both the NV bare resonance ($|0\rangle \rightarrow |-1\rangle)$ and thermally populated magnons at the FMR enhanced by quadratic mixing with the coherent magnons induced by the stripline MW drive. (c) Dressed state single arm sequence reveals coherent interference between the probe field and propagating magnons in the YIG. Light blue lines in (b) and (c) are fits of $\langle\phi^{2}\rangle$ data to the model described in supplemental section D3.}
    \label{fig:4}
\end{figure}

A 50\,$\mu$m $\times$ 50\,$\mu$m diamond chiplet containing a dense ensemble of NV centers is positioned approximately 300~nm above a 100~nm thick YIG thin film. A $\sim$10\,$\mu$m wide gold stripline patterned on the YIG surface both applies the probe field to the NV centers and excites spin waves in the YIG (Fig.~\ref{fig:4}(a)). A static external magnetic field along the NV axis is tuned such that the YIG ferromagnetic resonance (FMR) lies approximately 300~MHz above the NV $|0\rangle \to |-1\rangle$ transition used in this experiment (see supplemental section \ref{chap:characterisation_YIG}), ensuring that the qubit frequency is below the magnon gap. We highlight that this configuration is inaccessible to conventional NV DD noise magnetometry due to limitations on the Rabi rate.

In this experiment, the effective NV Rabi rate $\Omega_\text{eff}$ has three contributions: the direct stripline drive $\Omega_\text{direct}$, coherent magnons launched into the YIG $\Omega_\text{magnon}$, and the intrinsic thermal magnon noise $\Omega_\text{signal}$. As before, the AC Stark shift of the NV qubit depends on $\Omega_{\text{eff}}^{2}$:

\begin{equation}
    \Omega_\text{eff}^2= \Big(\Omega_\text{direct}(f)+\Omega_\text{magnon}(f)+ \Omega_\text{signal}(f)\Big)^2.
    \label{eq:YIG_eff_rabi}
\end{equation}
The choice of filter function $W_\tau$ in Eq.~\ref{eq:var_def_convolution} determines which term contributes to the phase variance.

We perform a dressed state spin echo measurement with $\tau$ = 500 ns and the probe field (i.e., direct drive field) applied during both arms of the pulse sequence (Fig.~\ref{fig:4}(b)), sweeping $f_\text{probe}$ from the NV resonance to 2.6~GHz and determining $\langle\phi^2\rangle$ at each frequency from NV decoherence measurements, as described in the previous section.

At $f_\text{probe} \approx 2.02$ GHz, a sharp peak in $\langle\phi^2\rangle$ appears from direct drive of the NV resonance due to the fact that the probe field, a Rabi drive when on resonance, has a random phase across repeated spin echo measurements. $\langle\phi^2\rangle$ then remains zero within measurement uncertainty until $f_{\text{probe}}$ reaches the YIG FMR near 2.3 GHz, where a second peak in $\langle\phi^2\rangle$ is observed. This signal originates from dressed state mixing of the probe field and the magnon response in the YIG film. An NV $T_{1}$ measurement with the probe field applied confirms that this peak does not originate from nonlinearities in the YIG response (see supplemental section \ref{chap:characterisation_YIG}).

To understand these features, we expand the square in Eq.~\ref{eq:YIG_eff_rabi} and drop $\Omega_\text{signal}^2$ as before. Two types of contributions remain: a coherent part $\Omega_\text{coherent}^2 \propto  (\Omega_\text{direct} + \Omega_\text{magnon})^2$ resulting from the direct drive (i.e., probe field) and the induced YIG magnons; and a cross term $\Omega_\text{mix}^2 \propto \Omega_\text{signal} \cdot (\Omega_\text{direct} + \Omega_\text{magnon})$ that mixes thermal magnon noise with the coherently enhanced probe field.

As with the test signal experiment described previously, the symmetric spin echo protocol cancels the deterministic coherent contribution, leaving only the cross term. The experimentally determined NV spin phase variance in the YIG experiment is therefore related to the noise spectral density $\mathcal{S}$ via a modified version of Eq.~\ref{eq:var_def_convolution} in which the effective probe field amplitude is enhanced by the coherent magnon response: $B_{\text{probe}}^{\text{eff}} \propto \Omega_{\text{direct}} + \Omega_\text{magnon}$. We model the thermal magnon noise spectral density, $\mathcal{S}$, as a $k$-integral over the Rayleigh-Jeans distribution, the magnon spectral density, and an effective filter in momentum space due to the finite separation between the NV ensemble and the YIG (see supplemental section~\ref{chap:YIG_fits})~\cite{du_control_2017}. Fitting of this model to the experimentally determined NV phase variance (light blue line in Fig.~\ref{fig:4}(b)) yields an NV-sample distance $d \approx 450$~nm, which is comparable to the anticipated distance of about $300$ nm (see supplemental section~\ref{chap:YIG_fits}).

We next perform a measurement of the NV spin dephasing with the probe field (i.e., direct drive field) applied only during the first arm of the spin echo sequence with length $\tau = 500$ ns (Fig.~\ref{fig:4}(c)). We observe a peak in phase variance $\langle\phi^{2}\rangle$ near the NV $|0\rangle \rightarrow |-1\rangle$ transition at 2.02~GHz; a gapped region where $\langle\phi^2\rangle$ remains zero within measurement uncertainty; and a steep rise in $\langle\phi^{2}\rangle$ when $f_{\text{probe}}$ exceeds 2.3~GHz, after which oscillatory features appear in $\langle\phi^{2}\rangle$ as a function of $f_{\text{probe}}$.

These oscillatory features can be understood as arising from the coherent contribution $\Omega_\text{coherent}^2$, which creates a static AC Stark shift that is cancelled in the symmetric spin echo measurement, but remains in the single arm measurement, with significant effect likely due to inhomogeneous broadening of the NV spin resonance within the confocal volume. $\Omega_\text{coherent}$ contains a sum of two terms: the direct magnetic field experienced by the NV center from the stripline, and the magnetic field stemming from magnons launched in the YIG film~\cite{zhou_magnon_2021}. The rectangular stripline launches magnons with a wavelength-dependent efficiency described by a sinc-shaped filter function in $k$-space~\cite{zhou_magnon_2021}, which imprints a periodic structure onto $\Omega_\text{coherent}^2$ as a function of probe frequency (see supplemental section~\ref{chap:characterisation_YIG}).

Assuming a linear magnon dispersion, the $\langle\phi^{2}\rangle$ data can be fit to our model (see supplemental section D3)), with results shown by the light blue line in Fig. 4(c). The periodicity of oscillations in the frequency spectrum can be converted to a magnon group velocity via the stripline width, yielding $v_\text{magnon} \approx 1450$~m/s, a value comparable to other literature results~\cite{bertelli_magnetic_2020}. The suppressed first oscillation in the $\langle\phi^{2}\rangle$ data relative to this model could be a result of a smearing of the magnon gap and reduced inhomogeneous broadening at low $k$.

\section{Discussion}

This work demonstrates that the dressed state AC Stark shift can be used to extend the frequency bandwidth of qubit noise magnetometry well beyond the Rabi rate limit of conventional dynamical decoupling (DD) measurement protocols. The quadratic dressed state mixing induced by a coherent probe tone and a noisy RF signal field, embedded in a symmetric DD sequence that cancels the probe-induced background, turns the qubit into a continuously tunable spectrum analyzer with center frequency set by the probe field. No change in magnetic field or hardware is required --- only the probe frequency is swept to realize a sensitive, broadband qubit-based spectrum analyzer with nanoscale spatial resolution.

The sensitivity of the dressed state DD protocol scales as $(\Omega_\text{probe}/\Delta)^2$ in the large detuning limit $\Delta \gg \Omega_\text{probe}$, where $\Delta$ is the detuning of the probe field from the qubit bare resonance and $\Omega_\text{probe}$ is the probe field Rabi rate. The measurement bandwidth (i.e., largest $\Delta$ that can be probed with reasonable sensitivity) is therefore set by the probe tone strength. For NV centers, Rabi rates up to $\sim 100$~MHz are experimentally attainable~\cite{jung_impedance-tuned_2025,fuchs_gigahertz_2009}, implying a feasible spectrum analyzer bandwidth of several GHz. For example, in our dressed state DD measurements of a YIG film, we leveragedcoherent magnons to amplify a thermal magnon signal hundreds of MHz away from the bare NV resonance. In principle, MW resonators could be used to apply directed MW probe fields to the NV sensor without disturbing the sensing volume.

More broadly, the dressed state DD protocol is naturally compatible with extensions to DD spectroscopy such as correlation spectroscopy~\cite{glenn_high-resolution_2018}, entanglement-enhanced sensing~\cite{liu_demonstration_2015, zhou_entanglement-enhanced_2025}, and quantum-computing-enhanced protocols~\cite{allen_quantum_2025}. The protocol is also platform-agnostic: any qubit with an AC Stark shift --- including superconducting qubits and trapped ions --- could implement this technique.

\section{Acknowledgments}

We gratefully acknowledge discussions with Hoang Le, Johannes Cremer, Elizabeth Park, Zui Tao and Ruolan Xue. This work
was supported by the U.S. Army Research Laboratory, under Contract No. W911NF2420143; the Laboratory for Physical Sciences Jumping Electron Quantum Fellowship Program under Award No. H9823022C0029; and the University of Maryland Quantum Technology Center. A. Y. is supported by the Gordon and Betty Moore Foundation through Grant GBMF 12762, by the U.S. Army Research Office (ARO) under grant number W911NF-22-1-0248, and by the Co-design Center for Quantum Advantage (C2QA), a National Quantum Information Science Research Center of the U.S. Department of Energy (DOE). N.M. was supported by an appointment to the Intelligence Community Postdoctoral Research Fellowship Program at Harvard University administered by Oak Ridge Institute for Science and Education. Diamond synthesis-related efforts were supported by the US Department of Energy, Office of Basic Energy Sciences, Materials Science and Engineering Division (N.D., F.J.H., D.D.A.).

\bibliography{final_citations}

\clearpage
\onecolumngrid

\begin{titlepage}
    \centering
    \vspace*{3cm}
    {\LARGE\bfseries Appendix\par} 
    \vspace{1.5cm}
    {\large A nanoscale magnetic spectrum analyzer based on qubit dressed states\par}
    \vfill
    \today
\end{titlepage}

\onecolumngrid
\appendix
\tableofcontents
\clearpage


\section{Experimental Details}

\subsection{Experimental setup}

\subsubsection{Optical}
\label{chap:optical_setup}
The excitation is done with a 532 nm, Cobolt Samba laser operated at 25 mW. Optical initialization and readout pulses are generated using an acousto-optic modulator (AOM, Isomet 1250C-848) in a double-pass configuration. Half- and quarter-wave plates set the beam polarization, and a single-mode fiber serves as a spatial filter. The optical power after the double pass AOM drops to around 5mW and to 2.5 mW after the fiber. 

Beam positioning and scanning in two dimensions are provided by a fast steering mirror (Newport FSM-300-01). The light is focused onto the diamond sample with a 50× microscope objective (Olympus MPLFLN50X, NA = 0.8). The optical power before insertion into the microscope is around 0.7 mW. A dichroic mirror (Semrock FF560-FDi01) reflects the green excitation beam while transmitting the red NV photoluminescence (PL). The red PL signal is then further cleaned by a long-pass filter (Semrock BLP01-594R-25). It is then detected with an avalanche photodiode (APD, Excelitas SPCM-AQRH-13). For sample inspection, a white-light source and camera are available and the corresponding mirrors are remotely insertable.

\begin{figure}
    \centering
    \includegraphics[width=0.8\linewidth]{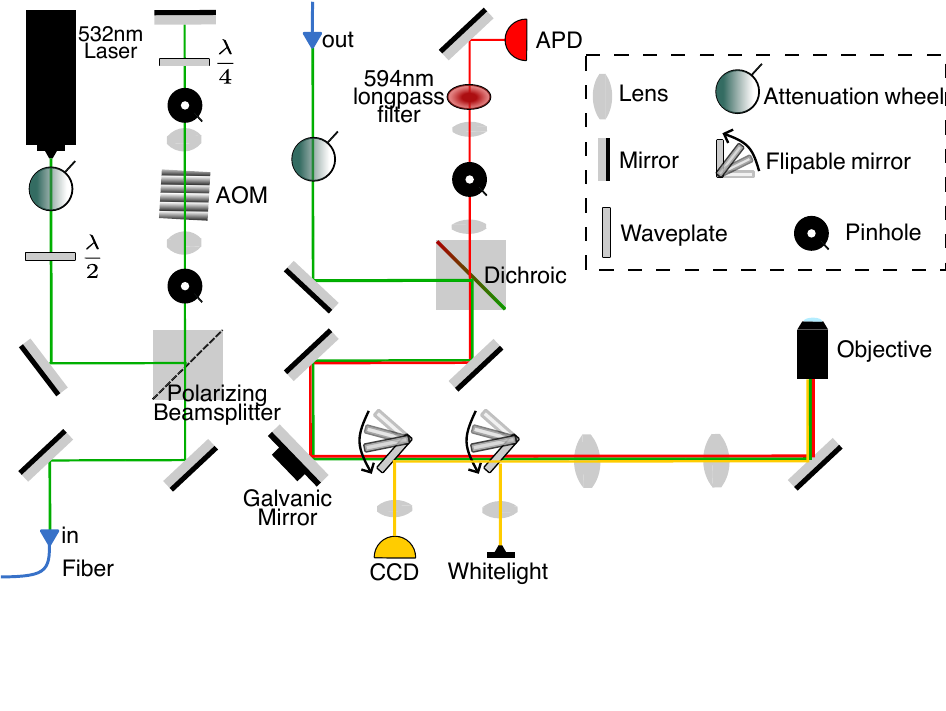}
    \caption{Scanning confocal microscope setup used for the NV characterization measurements.}
    \label{fig:optical_setup}
\end{figure}

\subsubsection{Microwave (MW)}
Two MW generators (R\&S SMB 100a) both serve as a LO for an IQ mixer (IQ1545LMP). The 4 I and Q channels are connected to an AWG (Tektronix AWG5014C) for the generation of RF noise signals with a specific noise spectra. The output of both mixers is then connected to a MW switch (Zaswa-2-50DRA+) which enables fast pulsing. Afterwards the outputs are combined using an (MiniCircuit ZB2PD-63-N+) and amplified (MiniCircuit ZHL-16W-43-S+). Delivery is done using an dangling wire mounted in close proximity to the NV. To quantifiy the MW signal a commercial spectrometer (Signalhound SA124B) is used before amplification.
\begin{figure}[h!]
    \centering
    \includegraphics[width=0.8\linewidth]{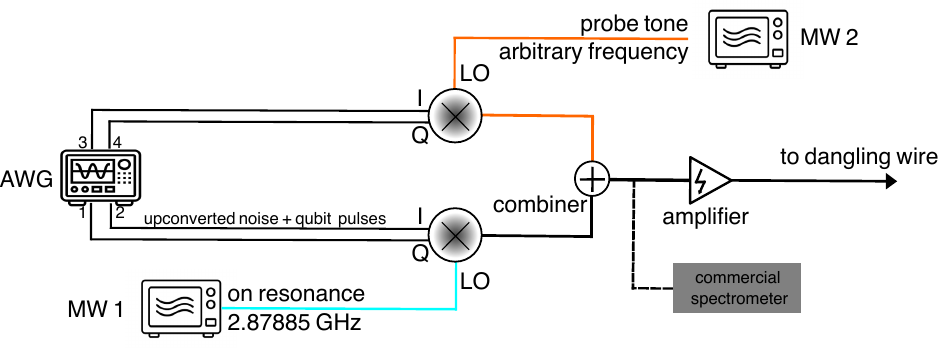}
    \caption{Microwave system used to generate signals for NV spin manipulation, signal and probe field generation and to drive the stripline. MW 1 manipulates the NV spin using the I and Q channels. For some experiments a noise signal is generated through the AWG and applied between the spin echo and ramsey pulses. MW 2 generates the probe field and can be swept in frequency.}
    \label{fig:mw_setup}
\end{figure}

\subsubsection{NV-Diamond pillar sample}
\label{chap:pillar_sample}

NV centers are created via delta-doping of nitrogen into an otherwise isotopically-purified (i.e. $\geq99.95\%$ $^{12}C$) [001] diamond grown via plasma enhanced chemical vapor deposition (PE-CVD). The nominal depth of the nitrogen dopants is 15 nm from the surface, with a nominal thickness of 5 nm and a nominal nitrogen doping of 0.5 ppm, followed with a buffer (distance from doped layer to the substrate) of 66.3 nm - all determined via SIMS based calibration on previous samples. The substrate is an electronic grade 2x2x0.5mm diamond from ElementSix that was fine-polished (CMP) to an Rq $<$ 0.3 nm and was strain released via ICP-RIE etching of $\sim$2.5 $\mu$m using a cyclic Ar/Cl and O plasma etch, as well as a 1200$^{o}C$ anneal for 2 h under vacuum $\leq 9 \times$10$^{-7}$ torr. Pre-growth the samples were reflux boiled in a sulfuric:nitric:perchloric acid mixture (1:1:1) followed by a heavy deionized water rinse to eliminate any surface contaminants. After CVD, the diamond was irradiated with 2MeV electrons with a fluence of 2$\times$10$^{14}$/cm$^{2}$. The diamond was then annealed under vacuum of approximately $\leq 9 \times$10$^{-7}$ torr at a temperature of 850$^{o}C$ for 4 hours, with a 850$^{o}C/min$ ramp and degassing plateau at 450$^{o}C$ for two hours, followed by another tri-acid reflux cleaning round.

Following NV creation and surface preparation, the nanopillar probes were patterned into the strain-etched face of the diamond membrane. The diamond was first bonded to a silicon carrier chip using Crystalbond mounting wax at 180$^{o}$C. A 10 nm titanium layer was then thermally evaporated onto the diamond surface; this thin metallic film simultaneously promotes adhesion of the subsequent e-beam resist and provides a conductive path to dissipate charge during electron beam lithography (EBL) on the otherwise insulating diamond substrate. The pillar etch mask was defined using hydrogen silsesquioxane (HSQ), a negative-tone e-beam resist in which exposed regions cross-link and remain after development. Three successive layers of Fox-16/HSQ were spun at 3000 RPM for 45 s each, with a 100$^{o}$C bake for 10 min between layers, yielding a total resist thickness of approximately 1 $\mu$m sufficient to provide the etch selectivity required for pillar formation. The pillar design -- a square lattice of circular disks $\sim$400--440 nm in diameter on a 7 $\mu$m pitch -- was then exposed onto the NV side of the membrane by EBL at a beam energy of 100 keV and an area dose of 5400 $\mu$C/cm$^{2}$. The exposed HSQ was developed in 25 wt.\% tetramethylammonium hydroxide (TMAH) for 30 s, after which the sample was rinsed by dipping sequentially through three consecutive cups of deionised (DI) water and finished with a final rinse in IPA. The pillar etch itself is a two-stage plasma process: first, an Ar/Cl$_{2}$ plasma removes the 10 nm titanium layer in the regions unprotected by the cross-linked HSQ mask (this chemistry does not attack the HSQ); immediately following, an O$_{2}$ plasma etches into the now-exposed diamond to a depth of $\sim$2 $\mu$m, forming the high-aspect-ratio nanopillars. Residual titanium and cross-linked HSQ remaining at the pillar tips were then stripped by a dip in hydrofluoric acid (HF). Finally, the diamond was released from the silicon carrier chip by dissolving the Crystalbond in acetone (or Remover PG), and the completed probes were subjected to a final tri-acid reflux clean matching that described above to ensure clean pillar surfaces prior to use. The pillar used has likely 2-3 NV centers and had a $T_2$ time around $10\mu$s.

\subsubsection{YIG sample}
\begin{figure}
    \centering
    \includegraphics[width=0.8\linewidth]{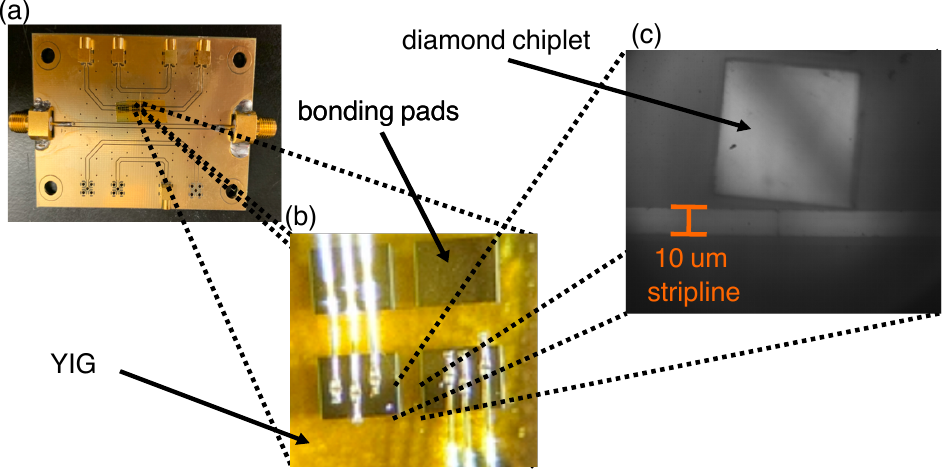}
    \caption{Visualization of MW deliverty for YIG experiment. (a) Full PCB with SMA and SMB connectors. (b)View of the 100nm thick comercial YIG film with patterned bonding pads. The 3 wire bonds per pad are clearly visible. (c)View of the $10$um stripline with the diamond chiplet containing the NV ensemble sitting on top of the YIG film.  }
    \label{fig:Yig_sample_zoom}
\end{figure}

For experiments on the YIG sample, MW delivery is through a $10$um gold stripline that is patterned by photolithography and connected to the PCB through wire bonds. A picture of the PCB, the diamond chiplet and the YIG sample is shown in Fig. \ref{fig:Yig_sample_zoom}. Typical values for $T_1$ in the chiplet are $0.46$ms and $T_2$ $2.3\mu$s.

\subsection{Noise creation}
\label{chap:Noise_creation}

To generate the RF noise signals in order to characterize the dressed state DD spectrum analyzer we use the fact that providing a RF signal at frequency $f_{IF}$ on both the I and Q channels of an IQ mixer will create two sidebands according to:
\begin{eqnarray}
&\cos\left(f_\mathrm{LO}t\right)\cdot\cos\left(f_\mathrm{IF}t\right)\\&=\frac{1}{2}\left[\cos\left(\left(f_\mathrm{LO}-f_\mathrm{IF}\right)t\right)+\cos\left(\left(f_\mathrm{LO}+f_\mathrm{IF}\right)t\right)\right]
\end{eqnarray}
where $f_{LO}$ is in our case the on resonant drive frequency of the NV spin transition. We use this simple technique, with IF signals generated by the AWG, to create arbitrary signal spectra.

To generate the broadband (rectangular) noise as seen in Fig. 2, we upload time traces to the AWG that follow Eq. \ref{eq:noise_sum}. Each trace is a superposition of many randomly phased cosine tones whose frequencies are uniformly distributed within a rectangular band around a chosen center frequency $f_\mathrm{noise}$.  
Uploading multiple traces ($\sim100$) and looping over them this creates a band-limited, approximately white noise signal centered at our frequency of choice. See Fig. 8 for an illustration.

The $i$-th noise time-trace is given by:
\begin{equation}
x_i(t)
= A_\mathrm{noise}
\frac{%
\displaystyle
\sum_{k=1}^{K}
\cos\!\bigl(2\pi f_k t + \phi_k\bigr)
}{%
\displaystyle
\max_t \left|\sum_{k=1}^{K}
\cos\!\bigl(2\pi f_k t + \phi_k\bigr)\right|
},
\label{eq:noise_sum}
\end{equation}
where
\begin{align*}
f_k &\sim \mathcal{U}(f_\mathrm{noise}-\tfrac{\mathrm{BW}}{2},\,f_\mathrm{noise}+\tfrac{\mathrm{BW}}{2}),\\
\phi_k &\sim \mathcal{U}(0,2\pi),\\
A_\mathrm{noise} \quad &\text{is the peak amplitude},\\
K \quad&\text{is the number of microtones (typically 512)},\\
\mathrm{BW} \quad&\text{is the total bandwidth},\\
f_\mathrm{noise}\quad &\text{is the center frequency of the noise band.}
\end{align*}

\begin{figure}
    \centering
    \includegraphics[width=1\linewidth]{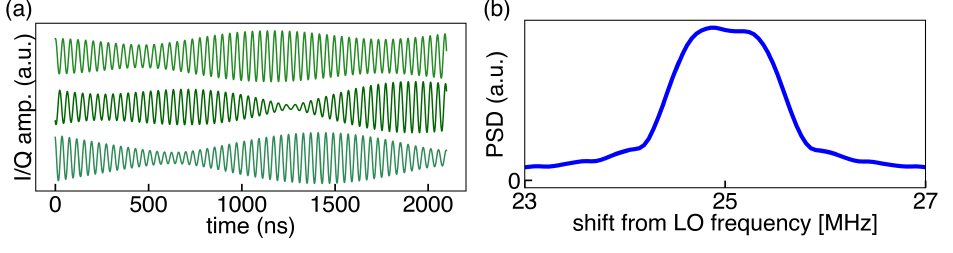}
    \caption{Visualisation of artificial noise time traces uploaded to the AWG. (a) Example of three different time traces uploaded $x_i(t)$ . The fast oscillations are at $f_{IF}$ while the envelope is given by the noise $BW$. (b) Averaged power spectral density (PSD) of the Fourier transformations of the sum of 100 different time traces.}
    \label{fig:Noise_creation_visualisation}
\end{figure}

\subsection{Measurement sequence}
\label{chap:measurement_sequence}

Initialization of NV centers to $m_s=0$ is done by applying a laser pulse of 2.1 $\mu$s length. We drive the NV electronic spin with MW pulses of frequency $f_{0\rightarrow +1}$ equal to the splitting between the $\vert 0\rangle$ and $\vert+1\rangle$ spin states. Applying a pulse with $f_{0\rightarrow +1}$ the for one quarter of the Rabi cycle duration rotates the NV spin from the state $|0\rangle$ to a coherent superposition, e.g., $|0\rangle + |+1\rangle /\sqrt{2}$. Typical Rabi rates of the setup are 2 MHz.

\begin{figure}
    \centering
    \includegraphics[width=1\linewidth]{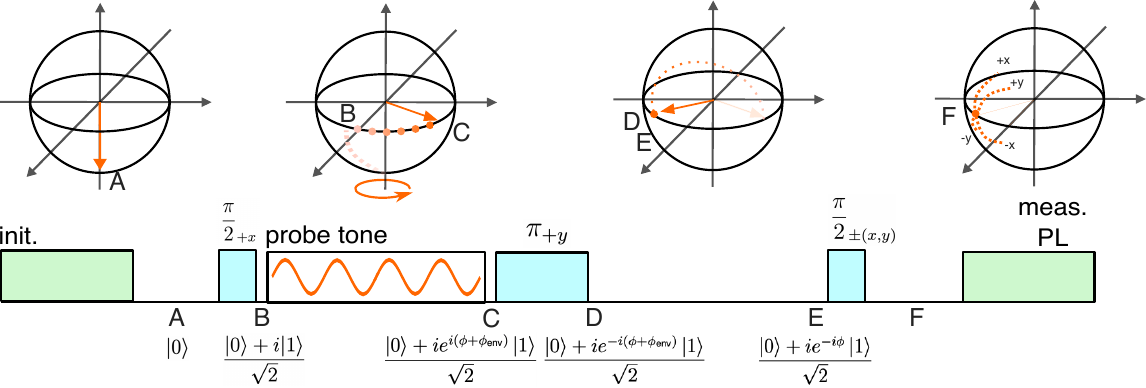}
    \caption{Visualization of spin echo sequence used to determine the NV spin phase and magnitude resulting from application of the probe field.}
    \label{fig:4axis}
\end{figure}

Fig.\ref{fig:4axis} shows details of the dressed state spin echo sequence used to aquire the experimental results shown in Fig. 1. After NV spin state initialization via a green laser (A),  a $\frac{\pi}{2}$ MW pulse prepares the NV spin into a coherent superposition in the transverse plane of the Bloch sphere (B). The probe field at frequency $f_\text{probe}$ is then applied for time $\tau$, inducing acquisition of phase $\phi$ (C). The full NV phase acquired, including the magnetic environment, is therefore $\phi+\phi_\text{env}$. The spin is then flipped $180^{\circ}$ by a $\pi$ MW pulse (D) and is left unperturbed for the same duration $\tau$ (E). This echo procedure cancels phase accumulation due to static magnetic fields from the environment, leaving a net accumulated phase of just $\phi$. Note that applying the probe field during the second half of the echo sequence we also cancel the effect of the AC stark shift on the NV phase as the AC stark shift acts as an effective, static magnetic field. 
As a final step, another $\frac{\pi}{2}$ MW pulse is applied and the NV spin projection onto the longitudinal (z) axis of the Bloch sphere is read out via NV PL induced by a green laser pulse. By iterating the rotation axis of the final $\frac{\pi}{2}$ MW pulse to be (+x, -x, +y, -y) through repeated runs of the experiment, we can reconstruct the NV spin’s position on the Bloch sphere prior to the final MW pulse (F). Defining $S_{\text{rot.\ axis}}$ as the PL measured after a rotation around the respective axis, we introduce the tomography signals
$\delta_{x/y} = S_{+x/y} - S_{-x/y}$.
From these quantities, the NV spin magnitude $\Lambda$ and accumulated phase $\phi$ prior to the final MW pulse follow as:
\begin{eqnarray}
\Lambda = \sqrt{\delta_x^2 + \delta_y^2},
\qquad
\phi = \arctan\left(\frac{\delta_y}{\delta_x}\right).
\label{eq:magnitude_and_phase_def}
\end{eqnarray}

\section{Theoretical framework}

\subsection{AC Stark shift and three-level dressed state picture}
\label{chap:derivation_Stark}
For a two-level system with an applied the Hamiltonian $\hat{H}$ can be written as:

\begin{eqnarray}
\hat{H} &=& h\,\gamma B_0 \hat{S}_z 
+ h\,\gamma B_\text{probe}\cos{(2\pi f_\text{probe}t+\phi)}\,\hat{S}_x.
\label{eq:def_H}
\end{eqnarray}

Let the eigenstates of $\hat{S}_z$ be $|0\rangle,|+1\rangle$; Eq.~\ref{eq:def_H} can then be expressed as:

\begin{eqnarray}
\hat{H} 
&=& 
\frac{h\,\gamma}{2}
\begin{pmatrix}
0 & B_\mathrm{probe}\cos(2\pi f_\mathrm{probe}t+\phi) \\
B_\mathrm{probe}\cos(2\pi f_\mathrm{probe}t+\phi) & 2 B_0
\end{pmatrix}.
\end{eqnarray}

The state $\Psi$ of the system evolves in time according to the Schrödinger equation:
\begin{eqnarray}
i\,h \frac{d\Psi}{dt} &=& 2\pi\, \hat{H}\, \Psi.
\end{eqnarray}

When we apply a unitary transformation $\hat{U}$ to $\Psi$ and transfer it to the rotating frame, 
with $\tilde{\Psi} = \hat{U}\Psi$, the transformed state $\tilde{\Psi}$ evolves according to 
the following effective Hamiltonian $\hat{H}'$:
\begin{eqnarray}
\hat{H}' &=& \hat{U}\hat{H}\hat{U}^\dagger + i\,\frac{h}{2\pi}\,(\partial_t \hat{U})\,\hat{U}^\dagger,
\end{eqnarray}
where
\begin{eqnarray}
\hat{U} &=& \exp\Bigl(i\,2\pi f_\mathrm{probe}\, t\, |+1\rangle\langle+1|\Bigr).
\end{eqnarray}

Using the rotating-wave approximation (RWA), $\hat{H}'$ can be written as:
\begin{eqnarray}
\hat{H}' &=& h
\begin{pmatrix}
0 & \dfrac{\Omega}{2}\, e^{i\phi} \\
\dfrac{\Omega}{2}\, e^{-i\phi} & \Delta
\end{pmatrix},
\end{eqnarray}
where 
\begin{eqnarray}
\Delta &=& f_\text{res} - f_\text{probe}, \qquad 
f_\text{res} = \gamma B_0, \qquad
\Omega = \gamma B_\text{probe}\,c \;=\; \frac{1}{T_\text{Rabi osc.}}.
\end{eqnarray}
Here $c$ is the matrix element $|\langle 0|\,\hat S_x/\hbar\,|1\rangle|$, which is $1/2$ for a spin $1/2$ system.
The eigenenergies $E_\pm$ of this effective Hamiltonian are then:
\begin{eqnarray}
E_\pm &=& \frac{h}{2}\,({\Delta \pm \sqrt{\Delta^2 + \Omega^2}}),
\end{eqnarray}
so that the energy splitting in the high detuning limit is given by:
\begin{eqnarray}
E_+ - E_- \;=\; h\,\sqrt{\Delta^2+\Omega^2}
\;\approx\; h\left(\Delta + \frac{\Omega^2}{2\Delta}\right)
\quad (\Delta\gg\Omega),
\label{eq:2lvl_energy_Taylor}
\end{eqnarray}
and the change in the splitting due to the probe field is therefore:
\begin{eqnarray}
\delta E \;=\; (E_+-E_-) - h \Delta \;=\; h\,\frac{\Omega^{2}}{2\Delta}.
\label{eq:energy_split_2lvl}
\end{eqnarray}

\subsubsection{Three level model}
\label{chap:three_lvl_model_theory}

The NV has a electronic spin $S=1$ system with basis $\{|0\rangle,|+1\rangle,|-1\rangle\}$. 
In the rotating frame at $f_\text{probe}$ and within the RWA, the effective Hamiltonian reads:
\begin{eqnarray}
\hat{H}' &=& h\,
\begin{pmatrix}
0 & \dfrac{\Omega}{{2}}\,e^{+i\phi_{+}} & \dfrac{\Omega}{{2}}\,e^{+i\phi_{-}} \\
\dfrac{\Omega}{{2}}\,e^{-i\phi_{+}} & \Delta_{+} & 0 \\
\dfrac{\Omega}{{2}}\,e^{-i\phi_{-}} & 0 & \Delta_{-}
\end{pmatrix},
\label{eq:H3_RWA}
\end{eqnarray}
where the detunings and Rabi rates are simply 
\begin{eqnarray}
\Delta_{+} = f_{0\rightarrow +1} - f_\text{probe}, \qquad
\Delta_{-} = f_{0\rightarrow -1} - f_\text{probe}, \qquad 
\Omega = \gamma B_\text{probe}\,c \;=\; \frac{1}{T_\text{Rabi osc.}}.
\end{eqnarray}
The  matrix element $|\langle 0|\,\hat S_x/\hbar\,|\pm1\rangle|$ is now $c=1/\sqrt{2}$ for both NV spin transitions.

The eigenvalues in the high detuning regime are in this case:
\begin{eqnarray}
E_{0} \approx -\,h\left(\frac{\Omega^{2}}{4\,\Delta_{+}}+\frac{\Omega^{2}}{4\,\Delta_{-}}\right), E_{+} \approx h\left(\Delta_{+}+\frac{\Omega^{2}}{4\,\Delta_{+}}\right),
E_{-} \approx h\left(\Delta_{-}+\frac{\Omega^{2}}{4\,\Delta_{-}}\right).
\label{eq:3lvl_ACshifts}
\end{eqnarray}

As we are interested in the change in energy between only one transition ($0\rightarrow +1$ int the present experiments) the relevant energy shift is therefore:

\begin{eqnarray}
    \delta E = (E_+-E_0) - h \Delta_{+} = h (\frac{\Omega^2}{2\Delta_+}+\frac{\Omega^2}{4\Delta_-}) = h\frac{\Omega^2}{2\tilde{\Delta}^2} 
    \label{eq:result_3lvl_RWA}
\end{eqnarray}

When the probe field is above the $0\rightarrow +1$ transition it is possible to write $\Delta_- = \epsilon +\Delta_+ $ where $\epsilon$ is the NV Zeeman splitting due to the external magnetic bias field. 

As our model is meant to be useful over a broad range of frequencies, we also account for the situation where we can no longer assume  $\Delta \ll f_\text{probe},f_{0\rightarrow +1} $ and hence we can no longer assume the RWA. We follow the calculation of \cite{van_der_sar_nanometre-scale_2015} appendix where they used the Schrieffer Wolff formalism. From there we get:
\begin{equation}
\begin{aligned}
\widetilde{\hat{H}}_{\text{eff}} 
= \hat{H}_0 
+ h
\begin{pmatrix}
A & 0 & 0 \\
0 & B - A & 0 \\
0 & 0 & -B
\end{pmatrix},
\end{aligned}
\end{equation}

\begin{equation}
A = \frac{\Omega^2 f_{0\rightarrow +1}}{2(f_{0\rightarrow +1} - f_\text{probe})(f_{0\rightarrow +1} + f_\text{probe})},
\quad
B = \frac{-\Omega^2 f_{0\rightarrow -1}}{2(f_{0\rightarrow -1} - f_\text{probe})(f_{0\rightarrow -1} + f_\text{probe})}.
\end{equation}
where the difference from \cite{van_der_sar_nanometre-scale_2015} is due to our definition of the Rabi rate to include the matrix element.

\begin{figure}
    \centering
    \includegraphics[width=0.5\linewidth]{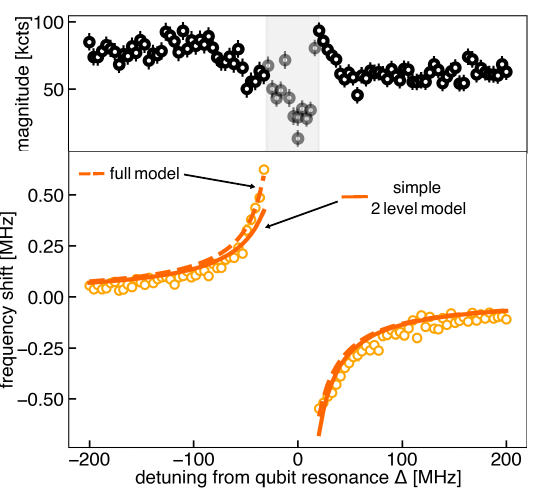}
    \caption{ Magnitude $\Lambda$ and frequency shift with respect to the detuning of the probe field. Close to resonance the magnitude decreases due to a resonant drive. The full and the simple model are fitted to the frequency shift.}
    \label{fig:full_model_fit}
\end{figure}

We note that in the limit of the RWA ($\Delta \ll f_\text{probe},f_{0\rightarrow +1} $) we can write $f_{0\rightarrow \pm1} + f_\text{probe}\approx2 f_{0\rightarrow \pm1} $ and recover the result of Eq. \ref{eq:result_3lvl_RWA}. As visible in \ref{fig:full_model_fit} the deviation between the full model and the simple 2 level approach is not  substantial especially in the high detuning regime. The phase data close to the resonance is not considered as the system begins to be driven on resonance (seen by the clear decrease in magnitude in the upper pannel).

\subsection{Qubit dressed state spectrum analyzer}
\label{chap:spectrum_analyzer}
The functionality of the qubit dressed state spectrum analyzer arises from four key physical effects or points:
\begin{enumerate}
    \item Applying an off resonant MW field to a two (or three) level system changes the energy difference of the eigenstates by $\delta E \approx \frac{\Omega_\text{}^2}{2\Delta}$.
    \item A relative shif in qubit eigenenergies leads to phase accumulation over time if the qubit is in a superposition state.
    \item Decoherence of a qubit follows the variance of the phase via $\Lambda(T)=\exp\!\left(-\frac{1}{2}\langle \phi^2(T)\rangle\right).$ \cite{machado_quantum_2023}
    \item A non linear mixing element, including the response of the qubit to RF fields arising from the AC stark shift enables downconversion of a noise spectrum by an arbitrary LO frequency.
\end{enumerate}
To connect these points, we introduce a situation in which the qubit experiences not only the coherent probe field but also an additional noisy signal field. In the rotating frame, these contributions add, such that the qubit evolves under an effective Rabi frequency
\begin{equation}
    \Omega_{\mathrm{eff}}(t)=\Omega_{\mathrm{probe}}(t)+\Omega_{\mathrm{s}}(t).
\end{equation}
Using \textbf{point~1}, the corresponding drive-induced shift of the dressed-state splitting relative to the bare states is
\begin{eqnarray}
    \delta E(t)/h
    &=& \frac{\Omega_{\mathrm{eff}}^{2}(t)}{2\Delta}
     = \frac{1}{2\Delta}\Big(\Omega_{\mathrm{probe}}^{2}(t)+\Omega_{\mathrm{s}}^{2}(t)
     +2\,\Omega_{\mathrm{probe}}(t)\Omega_{\mathrm{s}}(t)\Big)\nonumber\\
    &\approx& \frac{1}{2\Delta}\Big(\Omega_{\mathrm{probe}}^{2}(t)
     +2\,\Omega_{\mathrm{probe}}(t)\Omega_{\mathrm{s}}(t)\Big)\nonumber\\[3pt]
    &=& \frac{\gamma^{2}}{2c^{2}\Delta}\Big(B_{\mathrm{probe}}^{2}\cos^{2}(2\pi f_{\mathrm{probe}} t)
    +2 B_{\mathrm{probe}}\cos(2\pi f_{\mathrm{probe}} t)\,B_{\mathrm{s}}(t)\Big),
    \label{eq:full_noise_E_splitting}
\end{eqnarray}
where we neglect the term quadratic in the signal field, $\Omega_{\mathrm{s}}^{2}(t)\propto B_{\mathrm{s}}^{2}(t)$ as it is typically small and far off resonance.

From \textbf{point 2} we can derive the expected phase for one instance of the measurement through $h \phi = \int_0^\tau \delta E$. As we will average over thousands of individual experiments, this phase will not be deterministic over all experiments. We can define the variance of the accumulated phase due to the "noisy" AC Stark effect in analogy to \cite{machado_quantum_2023} :
\begin{eqnarray}
    h^2\langle\phi^2(\tau)\rangle = \int_0^\tau dt_1\int_0^\tau dt_2 \langle \delta E(t_1)\delta E(t_2)\rangle.
\label{eq:phisq_definition}
\end{eqnarray}

This is where \textbf{point 3} becomes crucial as the qubits phase variance can be determined by measuring the magnitude $\Lambda$ (Eq. \ref{eq:magnitude_and_phase_def}). In order to calculate this variance we can start by dropping the deterministic, noiseless part of Eq.~\ref{eq:full_noise_E_splitting}, $\delta E(t)_\text{det} /h =\frac{\gamma^2}{2c^2\Delta}B_\text{probe}^2\cos^2(2\pi f_\text{probe} t)$, as it will not contribute to the variance.
We are thus left with the \textit{crossterm} dependent on a multiplication of probe field and the signal field:
\begin{eqnarray}
\delta E_\text{mix}/h = \frac{\gamma^2}{c^2\Delta}B_\text{p}\cos(2\pi f_\text{probe} t)B_\text{s}(t).
\end{eqnarray}

The result enables the use of \textbf{point 4}. Multiplication of the noise signal and the coherent probe field, arising from the quadratic dependency of the AC stark shift on the effective magnetic field, effectively shifts upward the measurable frequency range of a qubit dynamical decoupling (DD) measurement sequence, enabling detection of the signal spectra.

The calculation starts by introducing the scaling prefactor
$\alpha \equiv \left(\frac{\,\gamma^2 B_\text{probe}}{c^2\Delta}\right)^2$ .
For an idealized two-level system, $\Delta$ is simply the probe detuning. For the NV center, a three-level treatment leads to the effective replacement
$\frac{1}{\Delta}\;\rightarrow\;\frac{1}{\Delta_+}+\frac{1}{2\Delta_-}$,
together with the corresponding choice $c=\sqrt{1/2}$, as discussed in ~\ref{chap:three_lvl_model_theory}.

We next calculate the argument of the integral in Eq.~\ref{eq:phisq_definition}:
\begin{eqnarray}
    \langle \delta E(t_1)\delta E(t_2)\rangle /\alpha
    \label{eq. AP_start_calculation}
    &=& \cos(2\pi f_\text{probe} t_1)\cos(2\pi f_\text{probe} t_2)\, \langle B_s(t_1) B_s(t_2)\rangle \\[3pt]
    &=& \cos(2\pi f_\text{probe} t_1)\cos(2\pi f_\text{probe} t_2)
    \int_{-\infty}^{\infty} df\,e^{-i2\pi f (t_2-t_1)}\, S_s(f).
\end{eqnarray}
Using $\cos x\cos y=\tfrac12[\cos(x-y)+\cos(x+y)]$ with
$x=2\pi f_\text{probe} t_1$ and $y=2\pi f_\text{probe} t_2$ yields:
\begin{equation}
    \langle \delta E(t_1)\delta E(t_2)\rangle /\alpha
    = \frac12\Big[\cos\!\big(2\pi f_\text{probe} (t_2-t_1)\big)+\cos\!\big(2\pi f_\text{probe} (t_2+t_1)\big)\Big]
    \int_{-\infty}^{\infty} df\,e^{-i2\pi f (t_2-t_1)}\, S_s(f).
\end{equation}

To justify neglecting the second term, it is convenient to switch to sum- and difference-time
coordinates:
\begin{equation}
    u=\frac{t_1+t_2}{2},\qquad v=t_2-t_1,
\end{equation}
for which $dt_1\,dt_2=du\,dv$ and the square $(t_1,t_2)\in[0,\tau]^2$ maps to
$v\in[-\tau,\tau]$ and $u\in[|v|/2,\ \tau-|v|/2]$.
In these variables:
\begin{equation}
    \cos\!\big(2\pi f_\text{probe}(t_2-t_1)\big)=\cos(2\pi f_\text{probe} v),
    \qquad
    \cos\!\big(2\pi f_\text{probe}(t_2+t_1)\big)=\cos(4\pi f_\text{probe} u).
\end{equation}
Since the correlator depends only on the time difference $v$, the sum-time contribution oscillates rapidly in $u$ and is therefore suppressed in the {double} integral:
\begin{equation}
    \left|\int_{|v|/2}^{\tau-|v|/2}\!du\,\cos(4\pi f_\text{probe} u)\right|
    =\left|\frac{\sin(4\pi f_\text{probe} u)}{4\pi f_\text{probe}}\Big|_{|v|/2}^{\tau-|v|/2}\right|
    \le \frac{1}{2\pi f_\text{probe}}
    \ll \tau \qquad (f_\text{probe}\tau\gg 1).
\end{equation}
Consequently, the term $\cos\!\big(2\pi f_\text{probe}(t_2+t_1)\big)$ is referred to as fast-rotating and
can be neglected compared to the surviving term, whose $u$-integration scales as $\int du \sim \tau$.

Keeping only the difference-time term results in:
\begin{eqnarray}
    \langle \delta E(t_1)\delta E(t_2)\rangle /\alpha
    &\simeq& \frac12\cos\!\big(2\pi f_\text{probe}(t_2-t_1)\big)
    \int_{-\infty}^{\infty} df\,e^{-i2\pi f (t_2-t_1)}\, S_s(f).
\end{eqnarray}
Now the "mixer" identity from \textbf{point 4} can be used:
\begin{eqnarray}
    2\cos\!\big(2\pi f_\text{probe}(t_2-t_1)\big)e^{-i2 \pi f (t_2-t_1)}
    = e^{-i2 \pi (f + f_\text{probe} )(t_2-t_1)}+e^{-i2 \pi( f - f_\text{probe} )(t_2-t_1)},
\end{eqnarray}
and the argument can be rewritten as
\begin{eqnarray}
    \langle \delta E(t_1)\delta E(t_2)\rangle /\alpha
    &\simeq& \frac14\int_{-\infty}^{\infty} df\,
    \Big(e^{-i2\pi(f + f_\text{probe} )(t_2-t_1)}+e^{-i2\pi(f - f_\text{probe} )(t_2-t_1)}\Big)\, S_s(f).
\end{eqnarray}

Inserting this back into Eq. \ref{eq:phisq_definition} to obtain the variance of the phase:

\begin{eqnarray}
    \langle\phi^2\rangle &=& 
    \frac{\alpha}{4}
    \int_0^\tau dt_1\int_0^\tau dt_2 
    \int_{-\infty}^{\infty} df\,
    \big(e^{-i2\pi(f + f_\text{probe} )(t_2-t_1)}+e^{-i2\pi(f - f_\text{probe} )(t_2-t_1)}\big)\, S_s(f) \\[3pt]
    &=& 
    \frac{\alpha}{4}
    \int_{-\infty}^{\infty} df\,S_s(f)\,
    \bigg(
    \Big|\int_0^\tau e^{i2\pi(f- f_\text{probe})t}\,dt\Big|^2 
    +  
    \Big|\int_0^\tau e^{i2\pi(f+ f_\text{probe})t}\,dt\Big|^2
    \bigg).
\end{eqnarray}

where we used that the only time dependence is in the exponentials,\\
$\big(\int_0^\tau e^{+i\theta t_1} dt_1\big)
\big(\int_0^\tau e^{-i \theta t_2} dt_2\big)
= \big|\int_0^\tau e^{i\theta t} dt\big|^2$.

As a last step we can now identify the scaled frequency filter function
\begin{eqnarray}
W_\tau(\theta) = \frac{\alpha}{2}\Big|\int_0^\tau e^{i\theta t}\,dt\Big|^2
\end{eqnarray}
and state:
\begin{eqnarray}
    \langle\phi^2\rangle(f_\text{p}) 
    &=& \frac{1}{2} \int_{-\infty}^{\infty} df\,
    S_s(f)\,
    \big[W_\tau(f - f_\text{probe}) + W_\tau(f + f_\text{probe})\big] \\[3pt] 
    &=& \int_{-\infty}^{\infty} df\,
    S_s(f) W_\tau(f + f_\text{probe}) \\[3pt] \label{eq:main_result_integral_from}
    &=& (S_s * W_\tau)(f_\text{probe})
    \label{eq:central_result_convo}
\end{eqnarray}
Here, it is assumed in the second equal sign that the noise is real and wide-sense stationary,which effectively means that it is symmetric $S(f) = S(-f)$ . \\

Eq. \ref{eq:central_result_convo} expression is the central result of this theoretical framework, as it captures the effective spectrum analyzer response in a compact form. The AC-Stark \textit{crossterm} acts as a frequency translator: by tuning the probe field frequency $f_\text{probe}$, the same intrinsic filter function $W_\tau$ is shifted along the frequency axis and therefore samples different parts of the noise spectrum. In other words, the measured phase variance $\langle\phi^2\rangle(f_\text{probe})$ is given by the convolution of the signal power spectral density $S_s(f)$ with a probe-frequency--shifted filter function. Mathematically this is equivalent to a down conversion of the noise.

\begin{figure}
    \centering
    \includegraphics[width=0.75\linewidth]{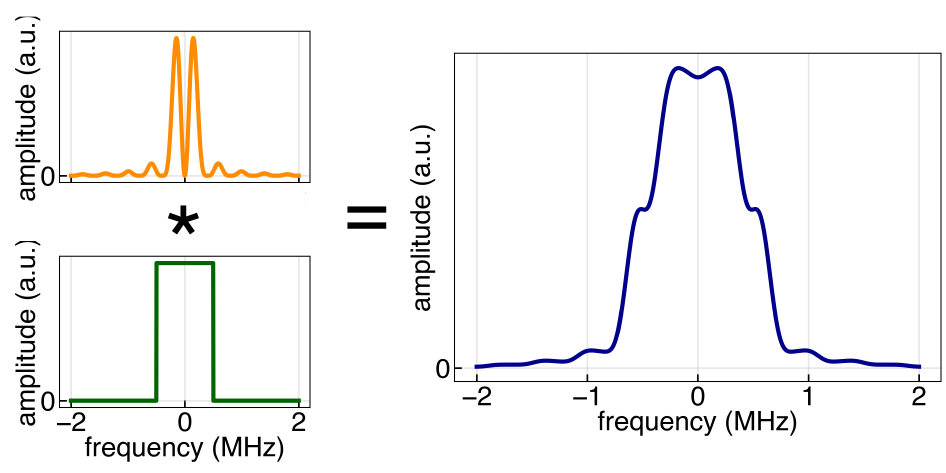}
    \caption{Visualization of the convolution of a spin echo filter function $W_{\tau=4000\text{ns}}(f)$ (orange) and a perfectly rectangular noise power spectral density $S_s(f)$ (green). The result is directly proportional to the probe frequency dependent phase variance $\langle\phi^2\rangle(f_\text{p})$ (blue). }
    \label{fig:simulation_convolution}
\end{figure}

Fig. \ref{fig:simulation_convolution} visualizes this result using the (rescaled) spin echo filter function:
\begin{eqnarray}
W_\tau(f)_\text{spin echo} = \frac{1}{2} \left(\frac{B_\text{probe}\gamma^2}{c^2\Delta}\right)^2
\frac{16}{(2\pi f)^2}\,
\sin^4\!\left(\frac{\pi f \tau}{2}\right).
\end{eqnarray}

\subsection{Scaling for broadband noise}
\label{chap:scaling_broad}

When the noise width of $\mathcal S_s(\omega)$ is much larger then the width of the filter function $W_\tau(\theta)$ ( $\approx1/\tau$) analysis of Eq. \ref{eq:central_result_convo} is straight-forward. In this case we can assume that the noise distribution is an effective constant $\mathcal{S}_s(f) = \mathcal{S}_s$ and calculate:
\begin{eqnarray}
    \langle\phi^2\rangle 
    &=&\frac{S_s}{2}\,\left(\frac{B_\text{probe}\gamma^2}{c^2 \Delta}\right)^2 \int_{-\infty}^{\infty}{df}\,
     \frac{16}{(2\pi f)^2}{\sin^4\frac{\pi f\tau}{2}} = \frac{S_s}{2}\,\left(\frac{B_\text{probe}\gamma^2}{c^2 \Delta}\right)^2  \tau \\
     &=& \beta  \frac{B_\text{probe}^2 \cdot \tau}{\Delta^2}  \quad\text{where}\quad \beta = 8S_s{ \,\gamma^4}{},
     \label{eq:result_ourapproach_coherence}
\end{eqnarray}
given that $c= \sqrt{1/2}$ for a three level system.

When comparing this result to a conventional DD measurement of qubit decoherenc, we observe an effective scaling with $\propto(\frac{\Omega_p}{\Delta})^2$, directly following from the rescaled filter function $W_\tau(f)$.

\section{Characterization results}

\subsection{Spectra}
\begin{figure}[h]
    \centering
    \includegraphics[width=1\linewidth]{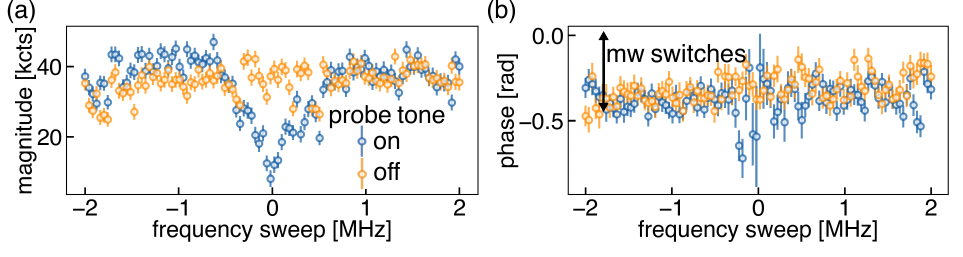}
    \caption{Dressed state spin echo and reference spin echo measurements for the probe field on and off, with application of a Lorentzian noise signal.  Data is expressed in terms of the unnormalized NV ensemble spin (a) magnitude $\Lambda$ and (b) phase $\phi$. For the dressed state spin echo measurement, mixing of the probe field and noise signal due to the AC Stark shift leads to a clear decrease in $\Lambda$ when the probe field frequency overlaps with the Lorentzian noise spectrum. No such decrease is observed for the reference measurement, as the noise signal is weak and detuned from the NV bare resonance by  $\Delta =25$ MHz. Here, the frequency zero on the x-axes is about $2.903$ GHz.  The MW switches induce an offset of about $-0.3$ rad in the NV spin phase determination, as indicated in (b). }
    \label{fig:raw_lorenzian}
\end{figure}
As the NVs in our experiments are also subject to environmental, noise we immediately follow each dressed state spin echo or Ramsey measurement with  a reference measurement without the probe field applied; and then take the logarithm of the ratio of the resulting NV spin magnitudes (with and without probe field)  to obtain the decoherence value. Fig. \ref{fig:raw_lorenzian} shows example data using the NV-diamond pillar for the probe field on and off, with application of a Lorenztian noise signal (as in Fig. 2b, lowest panel of the main text).  Figs. \ref{fig:raw_lorenzian}(a,b) show results for the (unnormalized) NV spin magnitude and phase, respectively. For the scaling in Fig. \ref{fig:2} the data was normalized to 1s measurement time. This was done by taking the time per measurement ($t_\text{measurement}=3700+540+2*4200$) and then obtaining a number for measurements per second ($N_{1s}=1/t_\text{measurement}$). This was then scaled with the actual taken measurements of $N_{total}=16$mil by $A_{scale} = \sqrt{N_{total}}/N_{1s}$.

\subsection{Experimental parameter calibrations and systematic dependencies}
\label{chap:scalings}

\subsubsection{Probe field amplitude and NV phase variance}
\begin{figure}[h]
    \centering
    \includegraphics[width=1\linewidth]{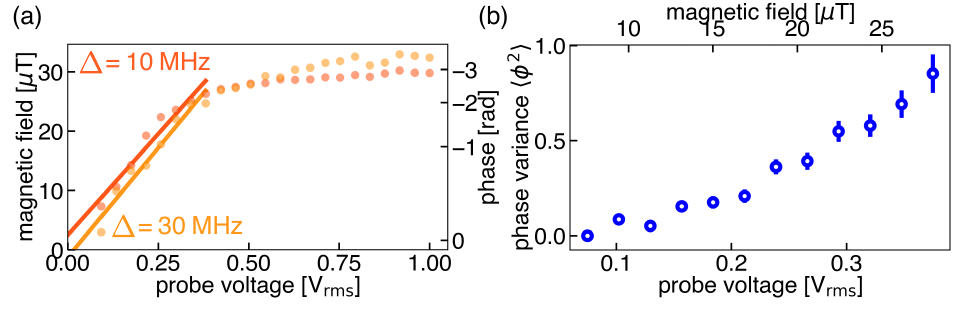}
    \caption{Calibration of the probe field. (a) $B_\text{probe}$ as a function of the voltage applied on the I and Q channels of the I-Q mixer and connected to MW2  to generate the probe field, for two values of probe field detuning $\Delta$ from the NV bare resonance. The output of the I-Q mixer saturates after the linear regime up to $0.4$ V. The magnetic field is calculated from the accumulated phase. (b) NV phase variance $\langle\phi^2\rangle$ as a function of the probe I-Q voltage and hence $B_\text{probe}$, determined from dressed state spin echo measurements in the presence of the rectangular noise spectrum shown in Fig. 2 of the main text. }
    \label{fig:power_sweep_explanation}
\end{figure}

Dressed state spin calibration measurements relates the voltage on the I-Q channels to the probe field magnetic field amplitude at the NV sensor. The calibration exploits the relation:
\begin{equation}
    B_\text{probe}=  \frac{{\sqrt{2}}}{\gamma}\Omega_\text{probe}=\frac{2}{\gamma} \sqrt{\frac{ |\phi| \Delta}{\tau}}  
    \label{eq:Probe_magnitude}
\end{equation}
which is a rewritten form of Eq. \ref{eq:coherent_phase} of the main text.

Example experimental results for this calibration are shown in Fig.~\ref{fig:power_sweep_explanation}(a), where $ B_\text{probe}$ (given by Eq. \ref{eq:Probe_magnitude} from experimentally determined values for $|\phi|$ and control parameters $\tau$ and $\Delta$) is plotted against the probe field voltage of the I-Q mixer. This calibration shows that the output of the I-Q mixer saturates at approximately $0.4~\mathrm{V}_{\mathrm{rms}}$, a value we independently confirmed using a commercial spectrum analyzer (Signalhound SA1W24B). For the experimental results presented in the main text and elsewhere in the supplement, we only use the linear regime up to $0.4~\mathrm{V}_{\mathrm{rms}}$.

To calibrate the induced NV phase variance $\langle \phi^{2} \rangle$ for a given voltage of the I-Q mixer (and hence probe field amplitude $B_\text{probe}$), we employ the rectangular noise spectrum shown in Fig. \ref{fig:2}, and with the probe field frequency tuned to be at the center of the noise band. Fig.~\ref{fig:power_sweep_explanation}(b) displays the experimentally determined phase variance $\langle \phi^{2} \rangle$ and the corresponding I-Q mixer voltage, as well as $B_\text{probe}$  at the NV center determined from the calibration obtained in panel (a).

Each data point displayed in Fig. \ref{fig:power_sweep_explanation} results from averaging over $2\cdot 10^{6}$experiment repetitions, implying that each of the $100$ noise time traces uploaded to the AWG is repeated (looped) about $2\cdot 10^{4}$ times per data point.



\subsubsection{$\tau$ dependence}


To investigate how the pulse sequence length influences the NV spin variance $\langle \phi^{2} \rangle$ we repeat the dressed state spin echo measurement for different echo-arm durations $\tau$ ranging from $0$ to $5000$~ns. To obtain the one-dimensional spectral representation in Fig. 3 of the main text, we define a noise signal region corresponding to the frequency band $f_k$ in Eq.~\ref{eq:noise_sum}, i.e., the frequencies at which noise is applied, with BW $= 1$ MHz. We then average all data points within this band and normalize the result by a reference region far detuned from the signal noise region. The reference region is defined as the frequency intervals in which no noise is applied, 
\begin{equation}
    f_\text{probe} \notin 
    \left[f_{\mathrm{noise}} - \frac{\mathrm{BW}}{2},\;
          f_{\mathrm{noise}} + \frac{\mathrm{BW}}{2}\right].
\end{equation}

\begin{figure}[h]
    \centering
    \includegraphics[width=0.55\linewidth]{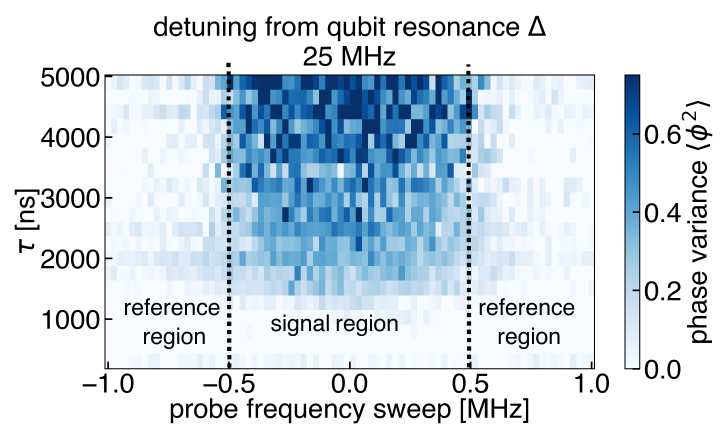}
    \caption{2D map of echo-arm duration $\tau$ dependency of the NV spin phase variance $\langle\phi^2\rangle$. Defined signal and reference regions are used to normalize the data to a 1D spectral representation, as in Fig. 2 of the main text..}
    \label{fig:tau_scaling}
\end{figure}

As shown in Fig. \ref{fig:tau_scaling}, the data largely follows the expected linear scaling of $\langle \phi^{2} \rangle$ with $\tau$. However, for $\tau < 100$~ns the measured phase variance is reduced. This deviation at short echo times results from the filter function becoming broader than the applied noise bandwidth of $1$~MHz, thereby no longer following the assumed limit of $S\gg W$ and thus decreasing the effective noise contribution.

\subsubsection{$\Delta$ dependence}

\begin{figure}[h]
    \centering
    \includegraphics[width=1\linewidth]{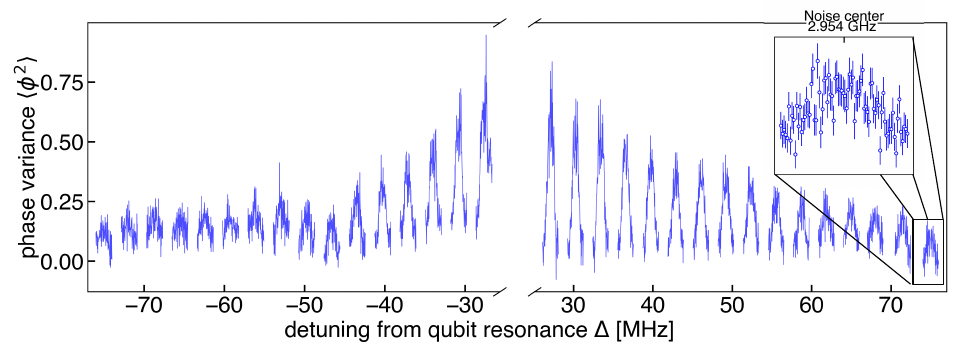}
     \caption{Dependence of the NV spin phase variance $\langle\phi^2\rangle$ dependency on the probe field detuning $\Delta$. For each value of the applied noise signal (with a rectangular spectrum), dressed state spin echo measurements are performed and the probe field frequency is swept as in Fig. \ref{fig:raw_lorenzian} is done. The spectrum of the applied noise signal is then shifted by $2.5$ MHz; and the dressed state spin echo measurements, with sweeping of the probe field frequency, are repeated. The Rabi rate $\Omega$ used for the measurements is $\sim2.5$ MHz, thus enabling operation in the large detuning regime, $\Omega \ll \Delta$.}
    \label{fig:detuning_scaling}
\end{figure}

Fig. \ref{fig:detuning_scaling} shows experimental characterization of the dependence of the qubit dressed state spectrum analyzer technique on probe field detuning $\Delta$ from the qubit bare resonance. The center frequency of an applied rectangular spectrum noise signal is iteratively shifted; and then NV dressed state spin echo measurements as a function of swept $f_\text{probe}$ are performed as above, in the large detuning limit, $\Delta \gg \Omega$. Similar to the characterization of $\tau$ dependence defined signal and reference regions are used to obtain the results in fig. 3 of the main text.

\newpage
\section{YIG}

\subsection{Damon-Eshbach magnons}

Consistent with the literature for similar samples we find that the most relevant magnons for our measurement are Damon-Eshbach modes within the YIG film.
The dispersion relation of Damon-Eshbach magnons in a film of thickness $l$ can be written, in the magnetostatic limit, as
\begin{equation}
f_{\mathrm{DE}}(k)
= \gamma \sqrt{\left[
B_\text{ext}\,
+ \frac{M_s}{2}
\right]^2
- \left(\frac{M_s}{2}\right)^2 e^{-2 k l}},
\label{eq:DE_dispersion}
\end{equation}
where $f_{\mathrm{DE}}(k)$ is the Damon-Eshbach magnon frequency, $k$ is the in-plane wavevector of the spin wave, $l$ is the thickness of the magnetic film, $B_\text{ext}$ is the magnitude of the external magnetic field component along the NV axis, and $M_s$ is the saturation magnetization \cite{du_control_2017}.

\begin{figure}
    \centering
    \includegraphics[width=0.8\linewidth]{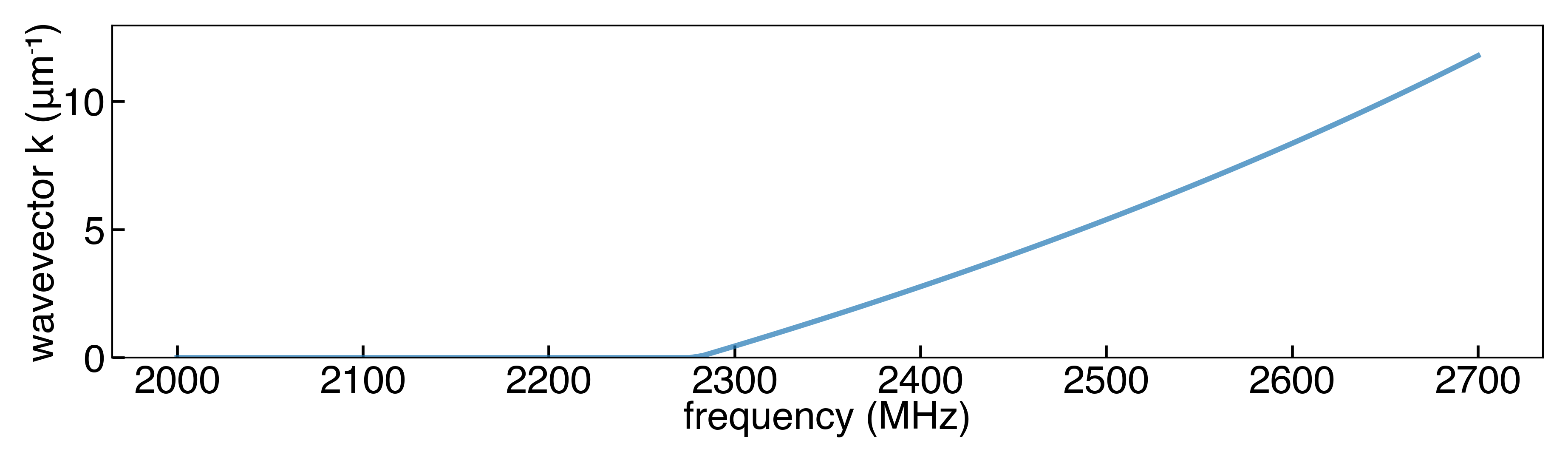}
    \caption{Dispersion relation of Damon-Eshbach magnons. The frequency range of the x-axis is scaled to the relevant frequency range for NV measurements reported here. The YIG film ferromagnetic resonance FMR is at $\approx 2280$ MHz for bias magnetic field $B_\text{ext}$ $303$ G.}
    \label{fig:magnon_dispersion}
\end{figure}

For the relevant experimental setup and YIG sample, relevant parameters are $B_\text{ext}=303$G, $M_s=1850$G film thickness is $100$ nm. For this configuration, varying $k$ traces out the Damon-Eshbach dispersion branch. This expected dispersion is displayed in Fig. \ref{fig:magnon_dispersion}, indicating that the FMR (i.e., $k\approx0$ magnon mode) is around $2280$ MHz, 

\subsection{Characterization by NV ODMR and T$_1$ measurements}
\label{chap:characterisation_YIG}

From previous NV measurements on a similar YIG film \cite{zhou_magnon_2021} $B_\text{ext} \approx 303$ G along the NV sensing axis is required for the FMR to have a higher frequency than the NV $f_{0\rightarrow-1}$ transition. This is crucial as the noise of the thermal magnons would otherwise directly decohere the NV spins. For this bias magnetic field, the NV bare $2020$ MHz while the magnon band begins around $2280$ MHz.

\begin{figure}[h]
    \centering
    \includegraphics[width=0.9\linewidth]{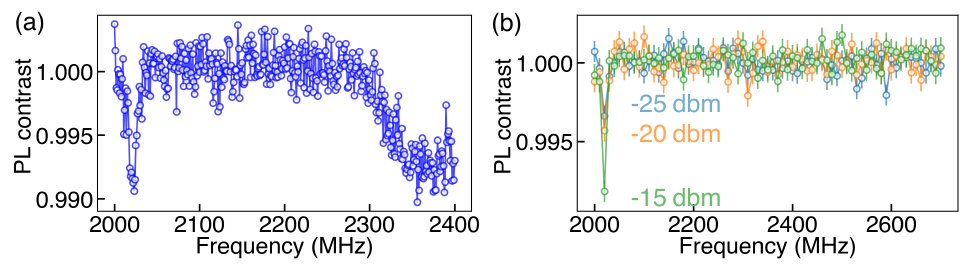}
    \caption{Characterization of effect of YIG film on NV PL behavior in presence of probe field drive by the stripline antenna. (a) Continuous NV ODMR measurement with probe field frequency iteratively swept over about 400 MHz, from the expected NV resonance to the FMR of the YIG film. The applied probe field power is $-20$ dBm.(b) Modified NV $\textrm{T}_1$ measurements for a timescale used in the dressed state spectrum analyzer pulse sequences ($500$ ns) rules out FMR- enhanced longitudinal NV spin relaxation as the origin of observed decreases in NV PL contrast and NV spin magnitude far from the NV bare resonance. Consistent results are found for three probe field powers. }
    \label{fig:YIG_checks}
\end{figure}

To check these assumptions continuous NV ODMR measurements are performed, with the probe field driven by the stripline antenna and swept in frequency from about 2000 to 2400 MHz. As seen in Fig. \ref{fig:YIG_checks}(a), the NV bare resonance is observed $2020$ MHz via a sharp decrease in NV PL contrast, consistent with direct drive of this transition by the probe field generated by the stripline antenna. A second, broad ODMR feature (reduction in NV PL contrast) is observed around $2275$ MHz consistent with the stripline driving magnons in the YIG film, which then scatter and drive the $0 \rightarrow+1$ transition. The frequency onset of this FMR is consistent with the predicted dispersion relation, see Fig. \ref{fig:magnon_dispersion}. The power to the stripline antenna is $-20$ dbm.

It is also important to test whether the observed decrease in NV PL contrast has an impact on timescales relevant for dressed state spectrum analyzer YIG measurements. This test is crucial to rule out that the observed decrease in NV spin magnitude $\Lambda$ in a spectrum analyzer measurement arises from enhanced $\textrm{T}_1$ spin relaxation (e.g from resonant drive or FMR-induced noise at the NV resonance) instead of the assumed NV spin decoherence induced by the AC Stark shift mixterm, as manifested in changes in the NV spin phase variance $\langle\phi^2\rangle$. The test consists of a modified NV $\textrm{T}_1$ measurement consisting of (i) laser initialization and readout pulses, each 6 $\mu$s long and (ii)  two $\tau =500$ ns pulses of the probe field between the laser pulses, with a brief delay (110 ns) between the probe field pulses.  The frequency $f_\text{probe}$ of the probe field pulses is then iterated across a range of about 600 MHz and the modified $\textrm{T}_1$ measurement is repeated for each $f_\text{probe}$ value.  Finally, the entire suite of measurements is repeated for three probe field powers (-25 dBm, -20 dBm, and -15 dBm). The 500 ns duration of the two probe field pulses is chosen to be consistent with the typical pulse durations used in the YIG experiments reported in the main text.  The result of this modified $\textrm{T}_1$ test, shown in Fig.\ref{fig:YIG_checks}(b) confirms that FMR-enhanced NV $\textrm{T}_1$ relaxation is insignificant on the timescales of our YIG measurements.

\subsection{Fits}
\label{chap:YIG_fits}

\paragraph{Spin Echo}

For the fit of the magnon spectra measured using the dressed state spin echo sequence, the following equation is used:
\begin{eqnarray}
    &&\mathcal{S}_s(f) \approx \!\!\int \mathrm{d}\mathbf{k}\cdot
    n(f,\mu)\, F(k,d)\, D(f,k)
    \approx  n(f,\mu) \int \mathrm{d}\mathbf{k'}\,\cdot\delta_{k'-k(f)}
    k'\, e^{-2 d k'} \left(1 - e^{-2 t_{\mathrm{yig}} k'}\right).
    \label{eq:momentum_filter}
\end{eqnarray}
Here $n(f,\mu)=\frac{k_B T}{h f-\mu}$ is the Rayleigh-Jeans distribution for which we estimate $\mu = 0$ as the probe field driving the magnons is relatively weak compared to past work and the fit is used for qualitative interpretation.
$F(k,d)$ is a filter function describing the field outside of the 2D plane. In this context $\mathrm{t}_{yig}$ is the thickness of the YIG film, in our case $100$ nm\cite{van_der_sar_nanometre-scale_2015}.
$D(f,k)$ denotes the magnon spectral density which we estimate here as $D(f,k)=\frac{1}{\pi}\,\frac{W}{W^{2}+\left[k(f)-k\right]^{2}}
\approx \delta_{k'-k(f)}$ given that the FMR spectrum is much narrower than that of the YIG thermal noise.  $k(f)$ is the dispersion relation of eq. \ref{eq:DE_dispersion}.

Inserting these relations into Eq. 3 of the main text gives the final fit function:

\begin{eqnarray}
    \langle\phi^2\rangle(f_\text{probe};A,d) = A \cdot \frac{1}{\Delta^2}\cdot \frac{k_B T}{h f_\text{probe}} \cdot
    k(f_\text{probe})^2\, e^{-2 d k(f_\text{probe})} \left(1 - e^{-2 t_{\mathrm{yig}}k(f_\text{probe})}\right).
\end{eqnarray}
where we assumed the limit broad noise limit discussed in supplemental section \ref{chap:scaling_broad} for the frequency filter and included all prefactors besides the detuning scaling $\frac{1}{\Delta^2}$ into the amplitude $A$. \textit{Note: the additional $k$ is due to the volume element of the 2D integral.}

\paragraph{Ramsey}

For the Ramsey measurements we use:
\begin{eqnarray}
    \langle\phi^2\rangle (f_\text{probe}) \nonumber &&\propto (B_\text{stripline}\chi(k,f))^2\\ &&\approx
    \!\!(\int \mathrm{d}\mathbf{k}\cdot L(k,d)\, D(f,k))^2\approx ( \int \mathrm{d}\mathbf{k'}\,\cdot\delta_{k'-k(f_\text{probe})}
    \,\text{sinc}(k'\cdot x_\text{stripline}/2))^2 
    \label{eq:ramsey_noise}
\end{eqnarray}

Here we introduce the stripline antenna filter function $ L(k,l)$. We approximate this filter by assuming that the antenna has a rectangular shape in the x-z plane. In that case we get $L(k,d)=\text{sinc}(k'\cdot x_\text{stripline}/2)$, as all the launched magnons under the stripline interfere and thereby effectively create a Fourier transform of the stripline. The final fit function is then:

\begin{eqnarray}
    \langle\phi^2\rangle(f;A, x_\text{stripline}) = A \cdot \frac{1}{\Delta^2}\cdot \sin(\pi \cdot k(f_\text{probe})\cdot x_\text{stripline}/2)^2,
\end{eqnarray}
where the $\pi$ originates from the sinc definition.
When assuming the linear dispersion relation $k\propto f$, we can additionally take a Fourier transform of the data from the oscillations of $\langle\phi^2\rangle$ as a function of probe frequency in the range $f_\text{probe}=[2400,2650]$ MHz (see Fig. \ref{fig:4} (c))and determine the oscillation spacing of  about $\approx 100$ MHz, which corresponds to a magnon group velocity $v_\text{magnon}\approx 1450\frac{\text{m}}{\text{s}}$.

\end{document}